\documentclass[amsmath,amssymb,aps,reprint,prl]{revtex4-2}
\usepackage{physics, amsmath}
\usepackage{graphicx}
\usepackage{float}
\usepackage{dcolumn}
\usepackage{bm}
\usepackage{hyperref}

\usepackage{xcolor}
\definecolor{deepblue}{HTML}{2D2F91}
\hypersetup{colorlinks=true,allcolors=deepblue}

\begin{document}
\preprint{APS/123-QED}
\title{\textbf{Coherent Control of Three-Level System Using Shaped Free Electrons}}
\author{Dixuan Wu$^{1,*}$}
\author{Jing Li$^{1}$}
\email{These authors contributed equally to this work.}
\author{Yuhan Jiang$^{1}$}
\author{Yunquan Liu$^{1,2}$}
\email{Yunquan.liu@pku.edu.cn}
\affiliation{%
$^1$State Key Laboratory for Mesoscopic Physics and Collaborative Innovation Center of Quantum Matter,\\
School of Physics, Peking University, Beijing 100871, China\\
$^2$Collaborative Innovation Center of Extreme Optics, Shanxi University, Taiyuan, Shanxi 030006, China\\
}
\date{\today}
\begin{abstract}
Three-level systems exhibit quantum interference effects absent in two-level systems, making them important for quantum optics. Here, we study the coherent interaction of a $\Lambda$-type three-level system with free electrons shaped by optical near fields. By treating the electron train as a quantum drive, we show that the interplay between electron modulation and the three-level system’s transition pathways induces tunable interference patterns. This interaction effectively realizes electron-mediated coherent population trapping (CPT). We identify a regime that enables complete population transfer between the two lower states and the preparation of a high-coherence superposition, manifested as dark states. In particular, these driven–dissipative steady states are independent of the initial state. Our work proposes shaped free electrons as a platform for steady-state coherent control of three-level systems, enabling atomic-scale state engineering.
\end{abstract}
\maketitle
Three-level systems are fundamental building blocks for coherent control in quantum optics~\cite{Scully_1997,Arimondo_1996,Fleischhauer_2005,Bergmann_1998,Vitanov_2017}, lasers~\cite{Silfvast_2004}, and quantum information processing~\cite{Hammerer_2010}, offering rich physics inaccessible in two-level systems. Their multiple transition pathways enable quantum interference effects, giving rise to coherent population trapping (CPT)~\cite{Arimondo_1976,Gray_1978,Arimondo_1996}, electromagnetically induced transparency~\cite{Harris_1990,Boller_1991,Field_1991,Fleischhauer_2005}, and stimulated Raman adiabatic passage~\cite{Kuklinski_1989,Schiemann_1993,Bergmann_1998,Vitanov_2017}. These mechanisms are central to quantum state engineering, including the preparation of dark states and superposition states that are robust against decoherence. However, conventional optical coherent control is fundamentally constrained by the diffraction limit~\cite{Born_1999}, which prevents individual addressing and coherent manipulation of quantum systems at the atomic scale, a capability essential for dense quantum information processing and nanophotonics~\cite{DiVincenzo_2000,Nielsen_2012,Bradac_2019}.

Recent breakthroughs in ultrafast transmission electron microscopy~\cite{Zewail_2010,Garc_a_de_Abajo_2021,Nabben_2023,Kuttruff_2024,LaGrange_2025} have propelled the development of free-electron quantum optics~\cite{ruimy2025,Shi2025,Garc_a_de_Abajo_2025}, an emerging framework that combines atomic-scale spatial resolution with ultrafast coherent control. An important part of free-electron quantum optics is photon-induced near-field electron microscopy (PINEM)~\cite{Barwick_2009,Park_2010,Feist_2015,kurman2021,yang2024}, in which optical near fields coherently shape free-electron wavefunctions into energy combs. Beyond electron--photon interactions, free-electron--bound-electron interactions constitute another important part of this field~\cite{ruimy2025,Shi2025,Garc_a_de_Abajo_2025}. The interaction between PINEM-modulated free electrons and two-level systems was first introduced in the framework of free-electron--bound-electron resonant interaction (FEBERI)~\cite{Gover_2020}, and later quantum treatments described the density-matrix dynamics of two-level systems driven by dilute trains of modulated electrons~\cite{Zhao_2021}. Related works have also developed shaped-electron protocols for atomic-scale quantum-state interrogation, preparation, and readout~\cite{Ruimy_2021,Zhang_2021,Zhang_2022,Abad_Arredondo_2025}. Despite these advances, the interplay between free electrons and multilevel systems remains largely unexplored. Specifically, how to use shaped free electrons to tailor quantum interference in three-level systems, especially to achieve population transfer and prepare coherent superposition states, remains an open question.

In this Letter, we investigate the resonant interaction between a train of PINEM-modulated free electrons and a $\Lambda$-type three-level system. Building on the framework of FEBERI in two-level systems~\cite{Zhao_2021}, we extend this treatment to a $\Lambda$-type three-level system, where the shaped free electrons couple to two optical transitions to enable atomic-scale steady-state coherent control. We combine a second-order $S$-matrix expansion for single-electron scattering with a Lindblad master equation for dissipation, treating the electron train as a periodic quantum driver acting on the dissipative three-level system. We map the steady-state diagram of the three-level system as a function of the PINEM coupling strength and the free-electron drift length, which reveals interference patterns governed by the interplay between the free-electron modulation and the transition pathways of the three-level system. Crucially, we identify a regime in which shaped electrons can (i) prepare a high-coherence superposition of the two lower states and (ii) realize complete population transfer between them. These outcomes are manifested as dark states in electron-mediated CPT. Our work shows that shaped electrons can serve as atomic-scale tools for driving quantum interference and engineering states in multilevel systems.

\begin{figure}
\centering
\includegraphics[width=1.0\columnwidth]{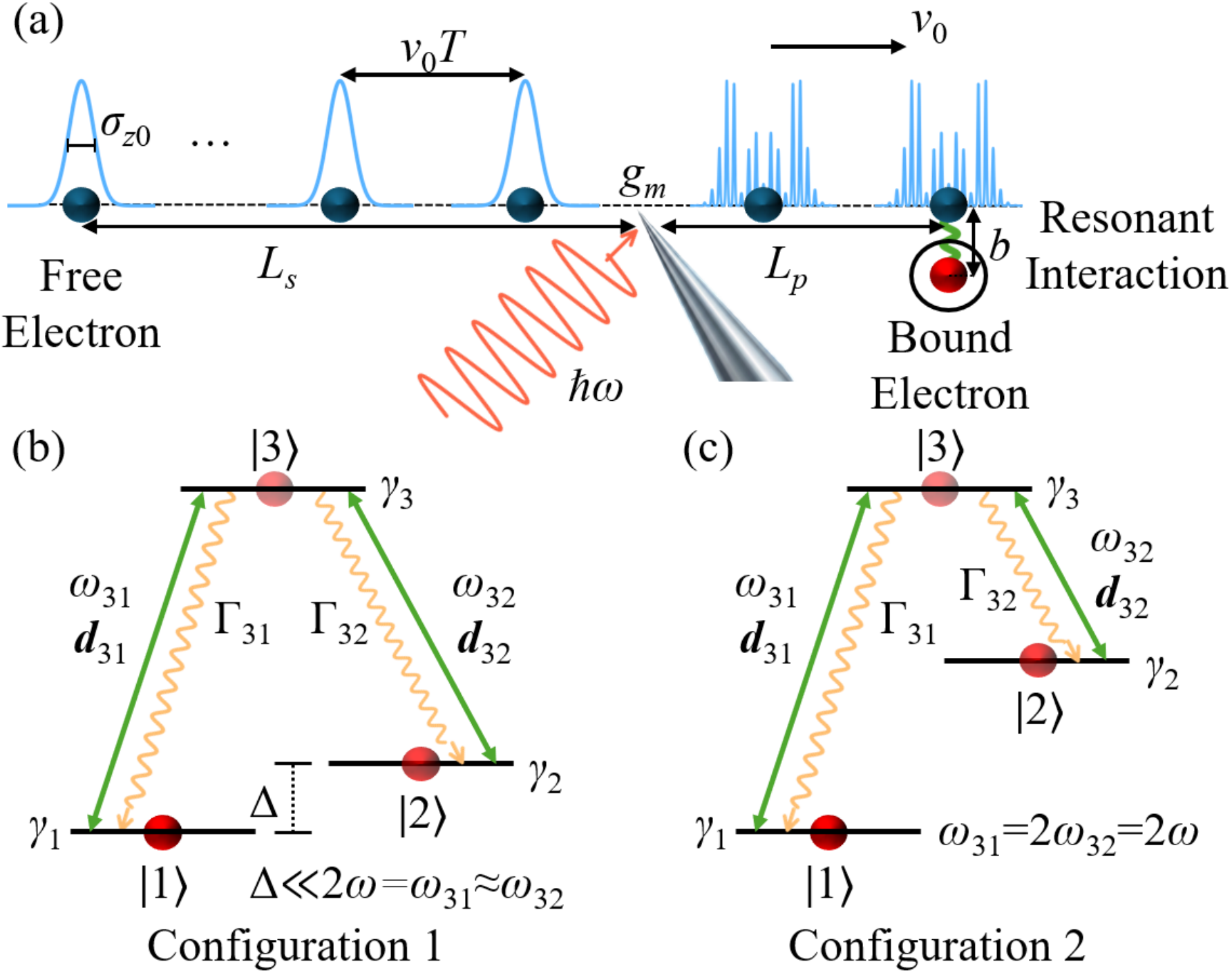}
\caption{Schematic of resonant interaction between free electrons and a $\Lambda$-type three-level system. (a) A train of free-electron wavepackets (velocity $v_0$, width $\sigma_{z0}$) is shaped via PINEM interaction with an optical near field (coupling strength $g_m$, frequency $\omega$) and drifts a distance $L_p$ before interacting with a bound electron at impact parameter $b$. (b),(c) Energy level diagrams for two considered configurations. Configuration 1 (b) involves a small detuning $\Delta \ll 2\omega = \omega_{31} \approx \omega_{32}$, whereas Configuration 2 (c) satisfies $\omega_{31} = 2\omega_{32} = 2\omega$. $\Gamma_{ij}$ and $\gamma_j$ denote spontaneous emission rates and pure dephasing, respectively.}
\label{fig:1}
\end{figure}

We consider the interaction of a train of shaped free electrons having a repetition period $T$ with a $\Lambda$-type three-level system, as depicted in Fig.~\ref{fig:1}(a). Each electron propagates with the group velocity $v_0$ and the central momentum $p_0=\gamma_0 m v_0$, where $\gamma_0=1/\sqrt{1-v_0^2/c^2}$. In terms of the wavevector deviation $q=(p-p_0)/\hbar$, the electron follows a Gaussian distribution in $q$-space with width $\sigma_q$, governed by the energy dispersion $E(q)\approx\gamma_0 m c^2+\hbar v_0 q+\hbar^2 q^2/(2\gamma_0^3 m)$. The electron initially undergoes free propagation over a distance $L_s$ before interacting with an optical near field of frequency $\omega$. The PINEM coupling strength is characterized by $g_m=\frac{e}{2\hbar\omega}\int E_z\left(z\right)\mathrm{e}^{-\mathrm{i}\omega z/v_0}\mathrm{d}z$, with $E_{z}$ denoting the longitudinal near-field component~\cite{Park_2010,Feist_2015,Shi2025,ruimy2025,Garc_a_de_Abajo_2025}. Following a subsequent drift of distance $L_p$, the longitudinal wavefunction in $q$-space prior to interaction with the three-level system is~\cite{Park_2010,Feist_2015,Zhao_2021}
\begin{align}
&\psi_e\left(q\right)=\left(2\pi\sigma_q^2\right)^{-\frac{1}{4}}
\mathrm{e}^{-\mathrm{i}\frac{E\left(q\right)L_p}{\hbar v_0}+\mathrm{i}\phi_0}
\sum_{n=-\infty}^{+\infty}\mathrm{J}_n\left(2\left|g_m\right|\right)\nonumber\\
&\exp\left[\mathrm{i}n\phi_{g_m}-\frac{\left(q-n\frac{\omega}{v_0}\right)^2}{4\sigma_q^2}-\mathrm{i}\frac{E\left(q-n\frac{\omega}{v_0}\right)L_s}{\hbar v_0}\right],\label{eq:wavefunction}
\end{align}
where $\phi_0$ is an initial phase, $\phi_{g_m} = \arg(g_m)$, and $\mathrm{J}_n$ denotes the Bessel function of the first kind of order $n$. In the calculation, we use the shaped electrons with the velocity of $v_0=0.5~c$, the wavepacket width of $\sigma_q=10^{6}$~m$^{-1}$ in $q$-space, and the period of $T=0.5$~ns. The PINEM modulation frequency is $\omega = 1.5 \times 10^{15}$~rad/s.

For the three-level systems shown in Figs.~\ref{fig:1}(b) and (c), the transition dipole moments are assumed to be parallel to each other and perpendicular to the electron trajectory, with constant magnitudes $|\boldsymbol{d}_{31}| = 2.16$~Debye and $|\boldsymbol{d}_{32}| = 2.03$~Debye for both configurations. We set the transition frequencies $\omega_{31} = 3.0 \times 10^{15}$~rad/s and $\omega_{32} = 2.9 \times 10^{15}$~rad/s for Configuration 1, while $\omega_{31} = 3.0 \times 10^{15}$~rad/s and $\omega_{32} = 1.5 \times 10^{15}$~rad/s for Configuration 2 to satisfy $\omega_{31}=2\omega_{32}=2\omega$. The spontaneous emission rates are determined by the dipole approximation $\Gamma_{3i} = \omega_{3i}^3 |\boldsymbol{d}_{3i}|^2 / (3\pi\varepsilon_0\hbar c^3)$ ($i=1,2$). The pure dephasing rates are set as $\gamma_1=0.95\times10^3$~s$^{-1}$, $\gamma_2=1\times10^3$~s$^{-1}$, and $\gamma_3=2\times10^6$~s$^{-1}$ for both configurations. The pre-modulation propagation length is $L_s=100~\mathrm{mm}$, and the impact parameter is $b=1$~nm.  

To describe the interaction between a free electron and a three-level system, we neglect spin~\cite{Zhang_2021}, retardation~\cite{Zhang_2021}, and exchange effects~\cite{Zhao_2021}, focusing solely on the Coulomb interaction treated within the dipole approximation. In the Schr\"odinger picture, the interaction Hamiltonian is given by
\begin{align*}
\hat{H}_I&=\sum_{\Delta q}\sum_{i,j} g_{ij}(\Delta q) \hat{b}_{\Delta q} \hat{\sigma}_{ij},\\
g_{ij}(\Delta q)
&=\frac{e}{2\pi\varepsilon_{0}L\gamma_{0}}
\left[|\Delta q|\mathrm{K}_{1}\left(\frac{|\Delta q|b}{\gamma_{0}}\right)\hat{\boldsymbol{e}}_{\perp}\right.\\
&\qquad\qquad\left.+\frac{\mathrm{i}\Delta q}{\gamma_{0}}
\mathrm{K}_{0}\left(\frac{|\Delta q|b}{\gamma_{0}}\right)\hat{\boldsymbol{e}}_{z}
\right]\cdot\boldsymbol{d}_{ij}.
\end{align*}
Here, $\mathrm{K}_{\nu}$ are modified Bessel functions of the second kind of order $\nu$, $b$ is the impact parameter, $\hat{\boldsymbol{e}}_{\perp}=\boldsymbol{b}/b$ and $\hat{\boldsymbol{e}}_{z}=(0,0,1)$ are unit vectors, $\boldsymbol{d}_{ij}$ is the transition dipole moment, and $L$ is the box-normalization length (see Section I in \cite{SM}). $\hat{b}_{\Delta q} = \sum_q \hat{c}_q^\dagger \hat{c}_{q+\Delta q}$ is the electron momentum transfer operator, and $\hat{\sigma}_{ij} = |i\rangle\langle j|$ is the transition operator of the three-level system. The indices $i,j$ are restricted to the allowed dipole transitions $\{13, 31, 23, 32\}$, with transition frequencies satisfying $\omega_{ij} = -\omega_{ji}$. The dynamics is governed by $\hat{S} = \mathcal{T} \exp[-\frac{\mathrm{i}}{\hbar}\int_{-\infty}^{+\infty} \mathrm{d}t \hat{H}_I(t)]$. The interaction Hamiltonian in the interaction picture is $\hat{H}_I(t) = \mathrm{e}^{\frac{\mathrm{i}}{\hbar}\hat{H}_0 t} \hat{H}_I \mathrm{e}^{-\frac{\mathrm{i}}{\hbar}\hat{H}_0 t}$, where $\hat{H}_0 = \sum_q E(q) \hat{c}_q^\dagger \hat{c}_q + \sum_i E_i\hat{\sigma}_{ii}$ represents the unperturbed Hamiltonian of the free electron and the three-level system. To quantify the interaction strength, we introduce the dimensionless coupling coefficient $G_{ij} = \frac{L}{\hbar v_0}g_{ij}\left(\frac{\omega_{ij}}{v_0}\right)$.
For the parameters here, we find $|G_{ij}| \approx 10^{-3} \ll 1$, which validates a perturbative expansion of the $S$-matrix up to the second order (see Section II in \cite{SM})
\begin{equation*}
\hat{S}=\hat{\mathbb{I}}-\mathrm{i}\sum_{i,j}G_{ij}\hat{b}_{\frac{\omega_{ij}}{v_0}}\hat{\sigma}_{ij}-\frac{1}{2}\sum_{i,j,k}G_{ik}G_{kj}\hat{b}_{\frac{\omega_{ik}}{v_0}}\hat{b}_{\frac{\omega_{kj}}{v_0}}\hat{\sigma}_{ik}\hat{\sigma}_{kj}.
\end{equation*}

The state of the three-level system after interaction with a single free electron, $\rho_b^{\text{(f)}}$, is obtained by
\begin{align}
\rho_b^{\text{(f)}} =\text{Tr}_e(\rho_{eb}^{\text{(f)}})= \text{Tr}_e (\hat{S} \rho_{eb}^{\text{(i)}} \hat{S}^\dagger)= \text{Tr}_e [ \hat{S} (\rho_e^{\text{(i)}} \otimes \rho_b^{\text{(i)}}) \hat{S}^\dagger],\nonumber
\end{align}
where $\rho_e^{\text{(i)}} = |\psi_e\rangle\langle\psi_e|$ is the initial state of the shaped free electron defined in Eq.~\eqref{eq:wavefunction}. The density-matrix changes of the three-level system, $\Delta \rho_{ij}=\langle i|(\rho_b^{\mathrm{(f)}}-\rho_b^{\mathrm{(i)}})|j\rangle$, can be derived analytically (see Section III in~\cite{SM}). We find that the PINEM modulation governs the interaction primarily through the free electron autocorrelation function
\begin{equation}
I(u) = \int_{-\infty}^{+\infty} \mathrm{d}q \, \psi_e(q+u)\psi_e^*(q).
\label{eq:4}
\end{equation}
Substituting Eq.~\eqref{eq:wavefunction} into Eq.~\eqref{eq:4}, we obtain an analytical expression for $I(u)$ that reveals the control mechanism:
\begin{align}
&I(u) = \mathrm{e}^{-\mathrm{i} L_p u} \sum_{l=-\infty}^{+\infty} \bigg\{ \mathrm{J}_l\left[ 4|g_m| \sin\left( \frac{\hbar \omega L_p u}{2\gamma_0^3 m v_0^2} \right) \right] \nonumber\\
&\times \mathrm{e}^{\mathrm{i}l(\phi_{g_m}+\frac{\pi}{2})+\mathrm{i}L_s\left(l\frac{\omega}{v_0}-u\right)}\mathcal{D}_l(u, L_s, L_p,\omega, v_0, \sigma_q)\bigg\},
\label{eq:Is_analytical}
\end{align}
where $\mathcal{D}_l$ is a Gaussian decay envelope that suppresses transitions for off-resonant momentum transfers (see Section III in \cite{SM}). By tuning the PINEM coupling strength $|g_m|$ and the drift length $L_p$, one can manipulate the argument of the Bessel function $\mathrm{J}_l$, thus controlling the magnitude of the transition amplitudes.

\begin{figure}
\centering
\includegraphics[width=1.0\columnwidth]{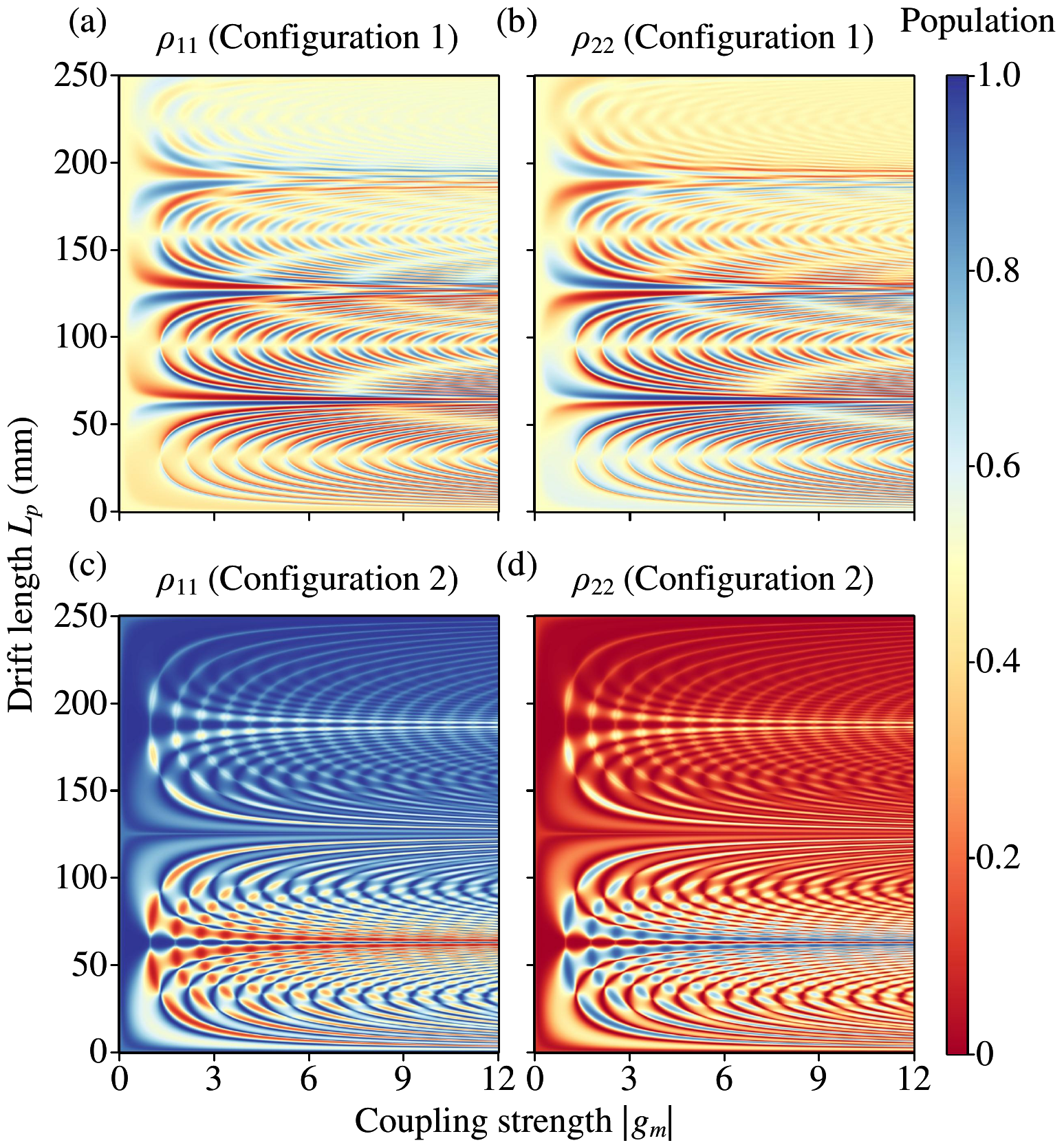}
\caption{Steady-state populations of the $\Lambda$-type three-level system driven by a shaped free-electron train. (a),(b) Two lower-state populations $\rho_{11}$ and $\rho_{22}$ for Configuration 1. (c),(d) Corresponding populations for Configuration 2. The populations are plotted as functions of the PINEM coupling strength $|g_m|$ and the free-electron drift length $L_p$, exhibiting distinct interference patterns governed by the interplay between electron energy modulation and two transition channels of the three-level system.}
\label{fig:2}
\end{figure}

Between successive electron interactions, the three-level system undergoes a dissipative evolution for a duration $T$, which is modeled by the Lindblad master equation \cite{rivas2012open,banerjee2018open}:
\begin{align}
\frac{\mathrm{d}\rho_b}{\mathrm{d}t}
=&-\frac{\mathrm{i}}{\hbar}\left[\hat{H}_{0b},\rho_b\right]+\sum_{i=1}^{2}\left(\hat{L}_{3i}\rho_b\hat{L}_{3i}^\dagger-\frac{1}{2}\left\{\hat{L}_{3i}^\dagger\hat{L}_{3i},\rho_b\right\}\right)\nonumber\\
&+\sum_{j=1}^{3}\left(\hat{L}_{\phi,j}\rho_b\hat{L}_{\phi,j}^\dagger-\frac{1}{2}\left\{\hat{L}_{\phi,j}^\dagger\hat{L}_{\phi,j},\rho_b\right\}\right),\label{eq:Lindblad}
\end{align}
where $\hat{L}_{3i} = \sqrt{\Gamma_{3i}} |i\rangle\langle 3|$ ($i=1,2$) describe spontaneous emission from $|3\rangle$ to $|i\rangle$, and $\hat{L}_{\phi,j} = \sqrt{2\gamma_j} |j\rangle\langle j|$ ($j=1,2,3$) account for pure dephasing. Integrating Eq.~\eqref{eq:Lindblad} over one period $T$ yields the inter-pulse evolution map. We use a phase-matched electron train with the accumulated phases over $T$ that satisfy $\omega_{31}T=2\pi m$ and $\omega_{32}T=2\pi n$ ($m,n\in\mathbb{Z}$). We define the total decay rate $\Gamma = \Gamma_{31} + \Gamma_{32}$ and the effective coherence decay rates $\gamma_{ij}$. Specifically, the coherence between $|3\rangle$ and $|i\rangle$ decays as $\gamma_{3i} = \Gamma/2 + \gamma_3 + \gamma_i$ ($i=1,2$), while the coherence between $|2\rangle$ and $|1\rangle$ decays as $\gamma_{21} = \gamma_2 + \gamma_1$.

To describe the cumulative dynamics induced by the electron train, we vectorize the density matrix of the three-level system into $\mathbf{v} = (\rho_{11}, \rho_{12}, \dots, \rho_{33})^\mathrm{T}$. The single-electron interaction acts as a linear map $\mathbf{v}' = (\mathbf{I} + \mathbf{D})\mathbf{v}$, where $\mathbf{D}$ accounts for the interaction-induced change in the density matrix (see Section IV, Eqs.~(70)--(97) in \cite{SM}). Subsequently, the dissipative evolution during one period $T$ is governed by a dissipation matrix $\mathbf{L}$, which accounts for spontaneous emission and pure dephasing. Combining these two processes, the state of the system after the $n$-th electron passes is calculated by the discrete map (see Section IV in \cite{SM}):
\begin{equation*}
\mathbf{v}^{(n)} = \mathbf{F} \mathbf{v}^{(n-1)} = \ldots = \mathbf{F}^n \mathbf{v}^{(0)},\quad \mathbf{F}=\left[\mathbf{L}\left(\mathbf{I} + \mathbf{D}\right)\right].
\end{equation*}

\begin{figure}
\centering
\includegraphics[width=1.0\columnwidth]{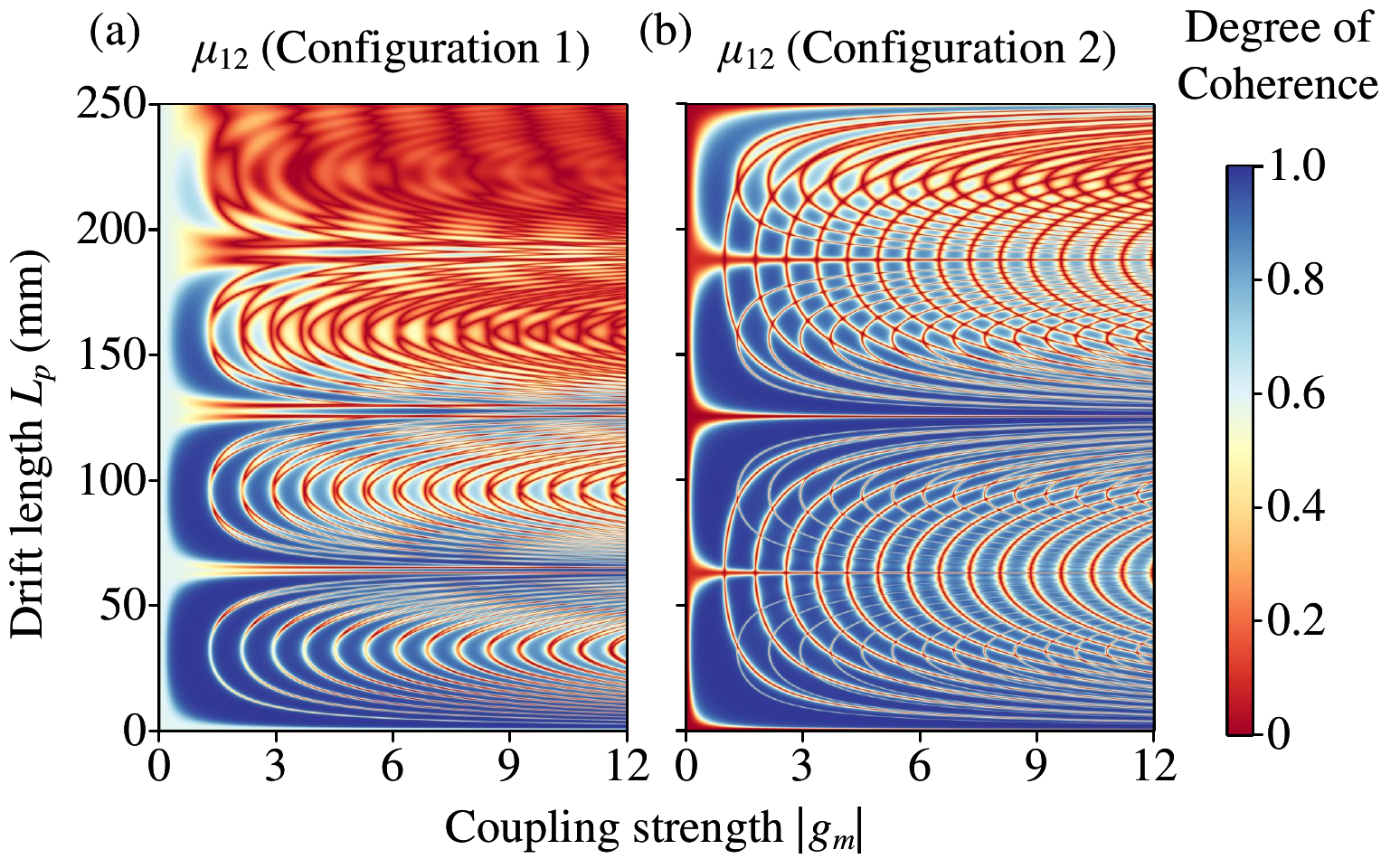}
\caption{Steady-state degree of coherence between the two lower states ($\mu_{12}$) induced by the electron train. The coherence maps are shown for (a) Configuration 1 and (b) Configuration 2 as functions of $|g_m|$ and $L_p$. High-coherence regimes correspond to regions of optimal population mixing.}
\label{fig:3}
\end{figure}

The steady-state solution $\mathbf{v}_{\text{ss}}$ is determined by the fixed-point condition $\mathbf{v}_{\text{ss}} = \mathbf{F} \mathbf{v}_{\text{ss}}$, subject to $\text{Tr}(\rho_b) = 1$. We can efficiently compute the steady-state populations and the degree of coherence presented in Figs.~\ref{fig:2} and \ref{fig:3} by solving the linear system equation $(\mathbf{F} - \mathbf{I})\mathbf{v}_{\text{ss}} = 0$ (see Section IV in \cite{SM}).

Figure~\ref{fig:2} illustrates the steady-state populations $\rho_{11}$ and $\rho_{22}$ as functions of the coupling strength $|g_m|$ and the free-electron drift length $L_p$. A prominent feature of these maps is the distinct periodic interference patterns, which originate from Eq.~\eqref{eq:Is_analytical}. Specifically, the modulation term $\sin[\hbar \omega L_p u / (2\gamma_0^3 m v_0^2)]$ in the Bessel functions defines a spatial periodicity of $z_u = 2\pi\gamma_0^3 m v_0^2 / (\hbar \omega u)$, where $u$ corresponds to the wavenumbers of the two transition channels, $\omega_{31}/v_0$ and $\omega_{32}/v_0$. Consequently, for Configuration 1 [Figs.~\ref{fig:2}(a) and (b)], the two sets of fringes nearly overlap and differ only by a slight offset along $L_p$, reflecting the weak detuning between $\omega_{31}$ and $\omega_{32}$. In contrast, for Configuration 2 [Figs.~\ref{fig:2}(c) and (d)], the fringe periodicities exhibit a strict 1:2 ratio, dictated by the condition $\omega_{31} = 2\omega_{32}$.

Moreover, the populations remain constant regardless of $L_p$ in the unmodulated limit ($|g_m| \approx 0$), indicating that unshaped free electrons are incapable of manipulating the population distribution between the two lower states. With stronger modulation, the population distribution becomes sensitive to both $|g_m|$ and $L_p$, manifesting itself as high-contrast interference fringes that define a versatile control landscape. The system can be driven into states where $\rho_{11} \approx 1, \rho_{22} \approx 0$ or conversely $\rho_{11} \approx 0, \rho_{22} \approx 1$, which validates effective population control. The interference fringes gradually blur with increasing $L_p$, diminishing the control authority. Shorter repetition periods $T$ and lower dissipation rates enhance both the maximum achievable population asymmetry and the effective $L_p$ range.

We quantify the coherence between the two lower states by the magnitude of the degree of coherence $\mu_{12} = |\rho_{12}|/\sqrt{\rho_{11}\rho_{22}}$, which satisfies $0\le\mu_{12}\le1$ and reaches unity for a pure superposition within the $\{|1\rangle,|2\rangle
\}$ manifold \cite{KarlBlum2012,Mandel_1995}. The resulting maps shown in Fig.~\ref{fig:3} exhibit periodic interference patterns and a gradual decay with increasing $L_p$, reflecting the features observed in Fig.~\ref{fig:2} and sharing the same physical origin rooted in Eq.~\eqref{eq:Is_analytical}. A significant finding is the generation of robust quantum coherence. Across the majority of the parameter space, $\mu_{12}$ remains close to unity, indicating the successful steady-state preparation of coherent superposition states between $|1\rangle$ and $|2\rangle$. A noticeable reduction in coherence occurs primarily in regions of extreme population imbalance, where the system is driven predominantly into a single basis state. Consistent with population dynamics, shorter repetition periods $T$ and weaker dissipation promote high-coherence superposition states and expand the range of $L_p$ over which clear interference fringes persist.

\begin{figure}
\centering
\includegraphics[width=1.0\columnwidth]{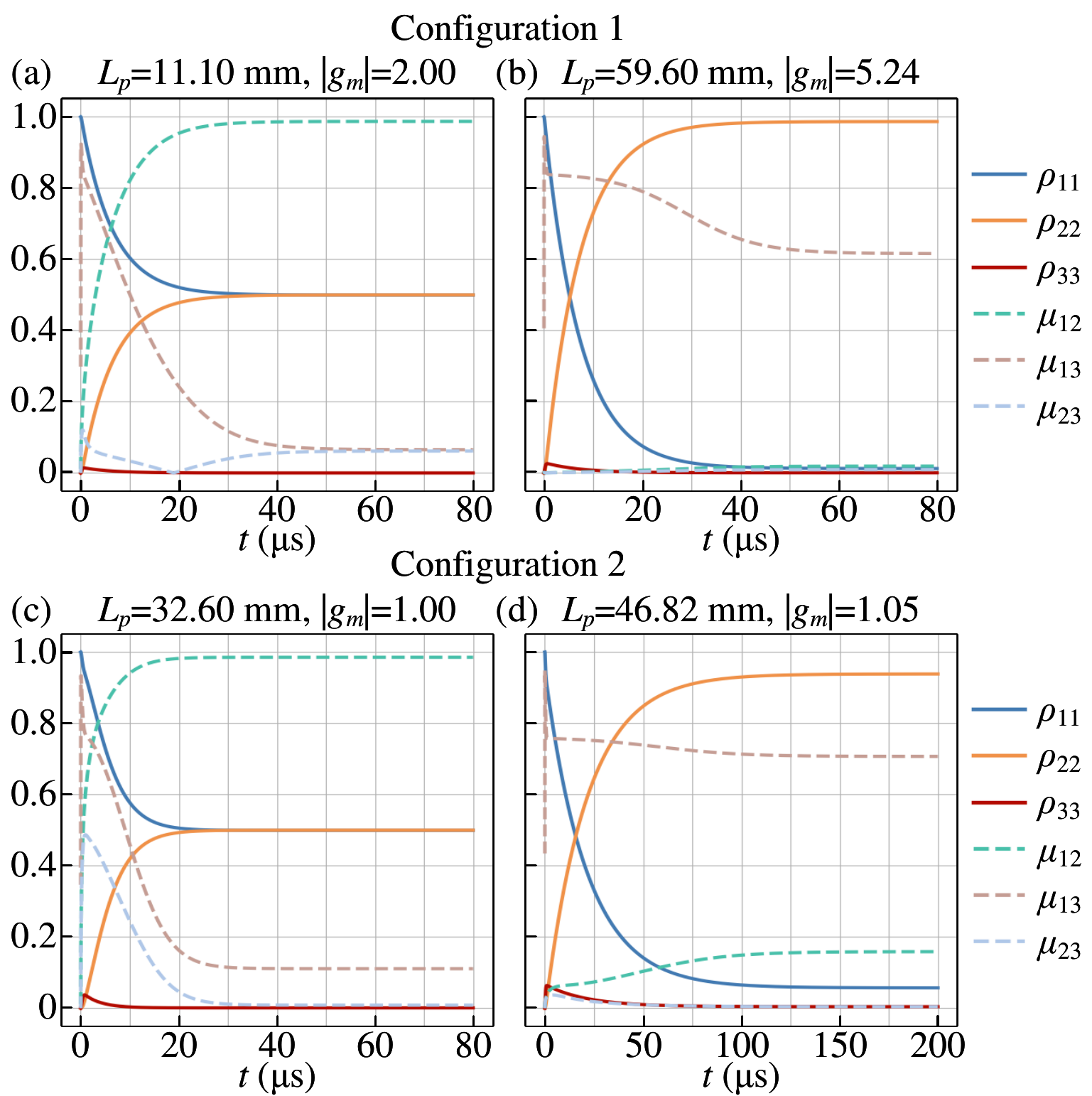}
\caption{Time evolution of the populations $\rho_{ii}$ and the degrees of coherence $\mu_{ij}$ for (a),(b) Configuration 1 and (c),(d) Configuration 2, with the initial state $\rho_{11}(0)=1$. (a),(c) Evolution toward a coherent superposition with balanced populations in $|1\rangle$ and $|2\rangle$. (b),(d) Realization of population transfer from $|1\rangle$ to $|2\rangle$. The different choices of $(L_p,|g_m|)$ for the two configurations reflect their distinct steady-state landscapes in the $(L_p,|g_m|)$ plane. Solid curves denote the populations $\rho_{ii}$, and dashed curves denote the degrees of coherence $\mu_{ij}$.}
\label{fig:4}
\end{figure}

The steady-state maps in Figs.~\ref{fig:2} and \ref{fig:3} reveal the global control landscape, identifying parameter regimes $(L_p, |g_m|)$ that support two primary control objectives for a $\Lambda$-type system: (i) a high-coherence superposition of the two lower states, and (ii) deterministic population transfer between the two lower states. To validate physical feasibility and determine the operational timescales for preparing these states, we study the temporal evolution of the system, as shown in Fig.~\ref{fig:4}. Starting from the initial state $\rho_{11}(0)=1$, the system converges to the predicted steady states within $\sim 10$--$100~\mu$s, corresponding to $\sim 10^{5}$ sequential electron-emitter interaction events. This confirms that the coherent drive provided by the shaped electron train can overcome dissipation and stabilize the system in desired quantum steady state.

For Configuration 1, a high-coherence equal-population superposition of $|1\rangle$ and $|2\rangle$ is obtained using the control parameters $(L_p,|g_m|)=(11.10~\mathrm{mm},2.00)$. As illustrated in Fig.~\ref{fig:4}(a), the system evolves to $\rho_{11}\approx \rho_{22}\approx 1/2$ with $\mu_{12}\to 1$. By retuning the parameters to $(59.60~\mathrm{mm},5.24)$, the system instead undergoes complete population transfer from $|1\rangle$ to $|2\rangle$, as depicted in Fig.~\ref{fig:4}(b).

The applicability of this control protocol extends naturally to Configuration 2, where the analogous target states can be realized with $(L_p,|g_m|)=(32.60~\mathrm{mm},1.00)$ for Fig.~\ref{fig:4}(c) and $(46.82~\mathrm{mm},1.05)$ for Fig.~\ref{fig:4}(d), respectively. Although the specific steady-state maps and convergence timescales vary between configurations, the underlying control mechanism remains robust and universal. For example, the system converges to the target state within $\sim50~\mu\mathrm{s}$ for Fig.~\ref{fig:4}(b) and $\sim150~\mu\mathrm{s}$ for Fig.~\ref{fig:4}(d). These variations stem from the distinct transition frequencies and parameter choices rather than a fundamental change in the physical principle.

These target states are non-equilibrium steady states maintained by the balance between the periodic electron drive and dissipation, and the generated coherence and population distribution persist as long as the electron train continues to interact with the three-level system. Furthermore, the final state is uniquely determined by the control parameters $(L_p,|g_m|)$ and is independent of the initial state of the three-level system, which enables deterministic state engineering without requiring precise knowledge of the pre-interaction history. Throughout the evolution, the excited state $|3\rangle$ remains only weakly populated: $\rho_{33}$ exhibits a small transient rise at early times and then decays to nearly zero. Since $|1\rangle$ and $|2\rangle$ are not directly coupled by a single transition, control relies on interfering two-step pathways mediated by $|3\rangle$ while keeping the lossy upper level largely unoccupied.

Such suppression of the lossy excited state is the hallmark of dark-state interference in laser-driven $\Lambda$-type three-level systems~\cite{Boller_1991,Fleischhauer_2005}. In CPT, two phase-coherent couplings on $|1\rangle\leftrightarrow|3\rangle$ and $|2\rangle\leftrightarrow|3\rangle$ define a dark superposition in the lower manifold with vanishing excitation amplitude to $|3\rangle$~\cite{Arimondo_1976,Arimondo_1996}. Under simultaneous driving and spontaneous decay of $|3\rangle$, repeated excitation–decay cycles transfer the population from bright components to the dark state~\cite{Gray_1978,Arimondo_1976}. Analogously, such a dark-state mechanism emerges in our framework through the discrete-time driven–dissipative protocol. By tuning $L_p$ and $|g_m|$, we engineer the free-electron autocorrelation function to tailor the complex quantum interference between the electron's energy modulation and the two transition pathways of the three-level system. Consequently, sequential electron scattering acts as a stroboscopic pump, driving the system toward a coherent steady state with suppressed excitation, which effectively establishes electron-mediated CPT. In this scheme, the requisite quantum coherence is carried by the electron wavefunction itself, obviating the need for two phase-locked optical tones directly to the emitter. This approach thus enables state engineering with atomic-scale spatial resolution beyond the optical diffraction limit. 

The robustness and experimental feasibility under non-ideal conditions are discussed in Section~V of \cite{SM}. Here we briefly summarize the main results. (i) For fluctuations of electron number per pulse, in the low-occupancy regime, the pulse-to-pulse electron number can be modeled by a Poisson distribution, and the results show that the target dynamics remain observable for mean occupancies $\lambda=0.15$ and $0.5$. (ii) For shot-to-shot fluctuations in $L_s$, $L_p$, and $\phi_{g_m}$, we include their effect only through Eq.~\eqref{eq:Is_analytical}, while possible changes in $T$ induced by $L_s$ and $L_p$ are excluded here. With this separation, the control remains robust for representative standard deviations $\sigma_{L_s}=1~\mu\mathrm{m}$, $\sigma_{L_p}=100~\mathrm{nm}$ in Configuration~1, $\sigma_{L_p}=50~\mathrm{nm}$ in Configuration~2, and $\sigma_{\phi_{g_m}}=0.1\pi$. (iii) For fluctuations of period $T$, the constraint is more restrictive because $T$ directly modifies the free evolution of the three-level system between successive electron interactions. The predicted effects survive for $\sigma_T=40$ attoseconds in Configuration~1 and $\sigma_T=5$ attoseconds in Configuration~2.  Although these requirements remain experimentally demanding, and related free-electron quantum-control schemes are still largely at the theoretical proposal stage~\cite{Gover_2020,Zhao_2021,Zhang_2021,Ruimy_2021,Zhang_2022,Abad_Arredondo_2025}, ongoing rapid progress in PINEM suggests a possible route toward future implementation~\cite{Feist_2015,Nabben_2023,Gaida_2024}.

In conclusion, we present a framework for coherent control of a $\Lambda$-type three-level system with PINEM-modulated free electrons. We show that the interplay between electron energy modulation and atomic transitions induces tunable steady-state interference, which realizes electron-mediated CPT. We identify a regime that achieves deterministic population transfer and prepares high-coherence superpositions of the two lower states, both of which correspond to dark-state control and are independent of the initial condition. A natural future extension involves introducing optical fields to drive a three-level system simultaneously, completing the interaction triad of free electrons, bound electrons, and light for synergistic hybrid control. Ultimately, this may offer avenues for mediating entanglement or implementing quantum logic gates through interactions between three-level systems and free electrons.\\

\textit{Acknowledgments}—This work was supported by the Natural Science Foundation of China (No.12334013 and No.124B2077), National Key R$\&$D Program (No.2022YFA1604301), and the Beijing Natural Science Foundation (No.QY25093 and No.QY25094).
\bibliography{mainref}
\end{document}